\documentstyle[aps,twocolumn,epsf]{revtex}

\begin{document}
\draft

\title{Observation of radiation pressure exerted by evanescent waves}

\author{D. Voigt, B.T. Wolschrijn, R. Jansen, N. Bhattacharya,
	R.J.C. Spreeuw, and H.B. van Linden van den Heuvell}

\address{Van der Waals-Zeeman Institute, University of Amsterdam, \\
         Valckenierstraat 65, 1018 XE Amsterdam, the Netherlands\\
         e-mail: spreeuw@wins.uva.nl}

\date{\today}
\maketitle

\begin{abstract}

We report a direct observation of radiation pressure, exerted on cold rubidium
atoms while bouncing on an evanescent-wave atom mirror. We analyze the
radiation pressure by imaging the motion of the atoms after the bounce.
The number of absorbed photons is measured for laser detunings ranging from
{190~MHz} to {1.4~GHz} and for angles from {0.9~mrad} to {24~mrad} above the
critical angle of total internal reflection. Depending on these settings,
we find velocity changes parallel with the mirror surface, ranging from
1 to {18~cm/s}. This corresponds to 2 to 31 photon recoils per atom.
These results are independent of the evanescent-wave optical power.

\end{abstract}


\section{Introduction}

An evanescent wave (EW) appears whenever an electromagnetic wave undergoes
total internal reflection at a dielectric interface.
The EW is characterized by an electric field amplitude that decays
exponentially with the distance to the interface.
The decay length is on the order of the (reduced) optical wavelength.
Cook and Hill \cite{CooHil82} proposed to use the EW as a mirror for slow
neutral atoms, based on the ``dipole force''.
EW mirrors have since become an important tool in atom optics
\cite{AdaSigMly94}.
They have been demonstrated for atomic beams at grazing incidence
\cite{BalLetOvc87} and for ultracold atoms at normal incidence
\cite{KasWeiChu90}.

Most experimental work so far, has been concentrated on the reflective
properties, {i.e.} the change of the atomic motion {\em perpendicular}
to the surface \cite{SeiAdaBal94}.
This is dominated by the dipole force due to the strong gradient of the
electric field amplitude.
In the present paper we report on our measurement of the force {\em parallel}
to the surface.
It has been mentioned already in the original proposal \cite{CooHil82}
that there should be such a force.
The propagating component of the wavevector leads to a spontaneous scattering
force, ``radiation pressure'' \cite{GorAsh80,Coo80}. 
We present here the first direct observation of radiation pressure exerted by
evanescent waves on cold atoms.
Previously, a force parallel to the surface has been observed for
micrometer-sized dielectric spheres \cite{KawSug92}.

In our experiment, we observe the trajectory of a cloud of cold rubidium atoms
falling and bouncing on a horizontal EW mirror. The radiation pressure appears
as a change in horizontal velocity during the bounce. We study the average
number of scattered photons per atom as a function of the detuning and angle of
incidence of the EW. The latter varies the ``steepness'' of the optical
potential.

Due to its short extension, on the order of the optical wavelength, an EW
mirror is a promising tool for efficient loading of low-dimensional optical
atom traps in the vicinity of the dielectric surface
\cite{SurfaceTraps,SprVoiWol}. In these schemes, spontaneous optical
transitions provide dissipation and cause the mirror to be ``inelastic'',
such that the final atomic phase-space density increases. This might open a
route towards quantum degenerate gases, which does not use evaporative cooling
\cite{KetDurSta99}, and could potentially yield ``atom lasers''
\cite{BECAtomLasers} which are open, driven systems out of thermal equilibrium,
similar to optical lasers. It is this application of EW mirrors which drives
our interest in experimental control of the photon scattering of bouncing
atoms.

This article is structured as follows. In {Sec.\,II} we summarize the
properties of the EW potential, including the photon scattering of a bouncing
atom. In {Sec.\,III} we describe our experimental setup and our observation of
radiation pressure. We investigate its dependence on the angle of incidence and
laser detuning. Finally, we discuss several systematic errors.

\section{The {evanescent-wave} atomic mirror}

\subsection{Optical dipole potential}

\begin{figure}[t]
  \centerline{\epsfxsize=8.0cm\epsffile{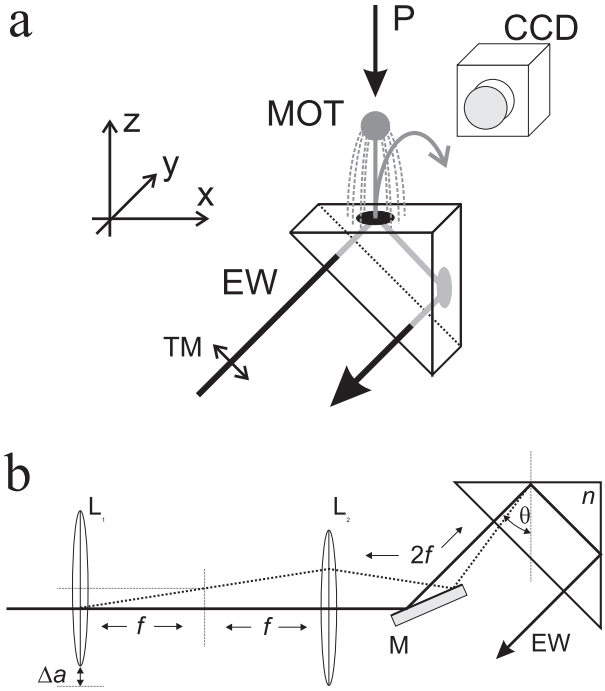}}
  \vspace*{0.5cm}
  \caption{Evanescent-wave atom mirror.
    (a)	Configuration in the rubidium vapor cell:
	magneto-optical trap (MOT), right-angle prism with refractive index $n$
	({6.6~mm} below the MOT, {gravity$\parallel$$z$}), evanescent-wave beam
	(EW), camera facing from the side (CCD, in $y$ direction),
	resonant fluorescence probe beam from above (P).
    (b)	Confocal relay telescope for adjusting the angle of incidence $\theta$.
	The lenses $L_{1,2}$ have equal focal length, {$f=75~$mm}.
	A translation of $L_1$ by a distance $\Delta a$ changes the angle of
	incidence by $\Delta\theta=\Delta a/{f n}$.
	The position of the EW spot remains fixed.
        M is a mirror.}
  \label{fig:SetupFigure}
\end{figure}

Our evanescent wave is created by total internal reflection at a glass surface
in vacuum \cite{Hec87}. We choose the $z$-direction as the surface normal and
the $xz$ plane as the plane of incidence (see {Fig.~\ref{fig:SetupFigure}}).
The EW can be written as
$E\left({\rm\bf r}\right)\propto\exp(i\,{\rm\bf k}\cdot{\rm\bf r})$,
where ${\rm\bf k}=(k_x, 0, i \kappa)$.
The wavevector has a propagating component along the surface,
$k_x=k_0 n \sin \theta$, where $k_0=2\pi/\lambda_0$ is the vacuum wave number,
$n$ is the refractive index and $\theta$ the angle of incidence.
Note, that $k_x>k_0$, since $\theta$ is larger than the critical angle
$\theta_c=\arcsin n^{-1}$.
The wavevector component perpendicular to the surface is imaginary,
with $\kappa=k_0\sqrt{n^2\sin ^{2}\theta -1}$.

The optical dipole potential for a {two-level} atom at a distance $z$ above the prism
can then be written as 	${\cal U}_{dip}(z)={\cal U}_0 \exp(-2\kappa z)$.
In the limit of large laser detuning, $\left|\delta\right|\gg\Gamma$,
and low saturation, $s_0\ll 1$, the maximum potential at the prism surface is
${\cal U}_0=\hbar\delta s_0$/2, with a saturation parameter 
$s_0\simeq(\Gamma/{2\delta})^2\,T I/{I_0}$ \cite{CohDupGry92}.
Here, {$I_0=1.65~$mW/cm$^2$} is the saturation intensity for rubidium and
{$\Gamma=2\pi\times 6.0~$MHz} is the natural linewidth. The intensity of the
incoming beam in the glass substrate is given as $I$. It is {\em enhanced} by
a factor $T$, that ranges between 5.4 and 6.0 for our TM polarized EW
\cite{FresnelCoefficients}. The detuning of the laser frequency $\omega_L$ with
respect to the atomic transition frequency
$\omega_0$ is defined as $\delta=\omega_L -\omega_0$. Thus a ``blue'' detuning,
$\delta>0$, yields an exponential potential barrier for incoming atoms.
Given an incident atom with momentum $p_i$, a classical turning point of the
motion exists if the barrier height exceeds the kinetic energy $p_i^2/2M$ of
the atom.

For a purely optical potential, the barrier height is given by ${\cal U}_0$.
In reality, the potential is also influenced by gravity and the attractive
{van der Waals} potential,
  \begin{equation}
     {\cal U}={\cal U}_{dip}+{\cal U}_{grav}+{\cal U}_{vdW}.
     \label{eq:TotalPotential}
  \end{equation}
The gravitational potential ${\cal U}_{grav}(z)\propto z$ can be neglected on
the length scale of the EW decay length $\xi\equiv 1/\kappa$. The van der Waals
potential ${\cal U}_{vdW}(z)\propto\left(k_0 z\right)^{-3}$ significantly
lowers the maximum potential close to the prism and thus decreases the
effective mirror surface on which atoms still can bounce. This effect has been
demonstrated by Landragin {\it et al.} \cite{LanCouLab96}.

\subsection{Photon scattering by bouncing atoms}

The photon scattering rate of a two-level atom in steady state at low
saturation can be written as
$\Gamma'=s\Gamma/2=(\Gamma/\hbar\delta)\,{\cal U}_{dip}$ \cite{CohDupGry92}.
An atom bouncing on an EW mirror sees a time dependent saturation
parameter $s(t)$.
Assuming that the excited state population follows adiabatically,
we can integrate the scattering rate along an atom's trajectory to get the
number of scattered photons,
  \begin{equation}
     N_{scat}=\int{\Gamma'(t)}dt
             =\frac{\Gamma}{\hbar\delta}
              \int_{-p_i}^{+p_i}{(\frac{{\cal U}_{dip}}{-\partial_z {\cal U}})}dp.
     \label{eq:PathIntegral}
  \end{equation}
For a purely optical potential, ${\cal U}\propto\exp(-2\kappa z)$, this leads
to an analytical solution:
  \begin{equation}
     N_{scat}=\frac{\Gamma}{\delta}\frac{p_i}{\hbar\kappa}.
     \label{eq:AnalyticalSolution}
  \end{equation}
The ``steepness'' of the
optical potential is determined by $\kappa$.
The steeper the potential,
the shorter the time an atom spends in the light field,
and the lower $N_{scat}$.
Note that $N_{scat}$ is independent of ${\cal U}_0$,
as a consequence of the $\exp(-2\kappa z)$ shape of the potential.
This is in fact the reason why a {two-level} description of the atoms is appropriate.
Strictly speaking, for a realistic atom
the potential strength depends on the magnetic sublevel $m_F$
through the {Clebsch-Gordan} coefficients.
For example, our {EW} is {TM}-polarized which,
close to the critical angle, is approximately linear.
Effectively, for every $m_F$ sublevel,
${\cal U}_0$ is then multiplied by the square of a Clebsch-Gordan coefficient,
ranging from 1/3 to 3/5 (for a $F=2\rightarrow F'=3$ transition).
Remarkably though,
the value of $N_{scat}$ remains unaffected and the same for all atoms.

One expects that an absorbed photon gives a recoil momentum $p_{rec}=\hbar k_x$
to the atom, directed along the propagating component of the EW. Experimentally
we observe this effect by the altered horizontal velocity of atom clouds after
the bounce. The spontaneous emission of photons leads to heating of the cloud
and thus to thermal expansion \cite{LanLabHen96,HenMolKaiA97}.

In principle, $N_{scat}$ is changed if other than optical forces are present.
For example, the van der Waals attraction tends to ``soften'' the potential
and thus to increase $N_{scat}$. We investigated this numerically and found it
to be negligible in our present experiment.

\section{The observation of radiation pressure}

\subsection{Experimental setup}

Our experiment is performed in a rubidium vapor cell. We trap about $10^{7}$
atoms of {$^{87}$Rb} in a magneto-optical trap (MOT) and subsequently cool them
in optical molasses to {10~$\mu$K}. The MOT is centered {6.6~mm} above the
horizontal surface of a right-angle BK7 prism
($n=1.51$, $\theta_c=41.4^{\circ}$ \cite{MellesGriotPrism}),
as shown in {Fig.\,\ref{fig:SetupFigure}a.

The EW beam emerges from a single-mode optical fiber, is collimated and
directed to the prism through a relay telescope
(see {Fig.\,\ref{fig:SetupFigure}b}).
The angle of incidence $\theta$ is controlled by the vertical displacement
$\Delta a$ of the first telescope lens, $L_1$.
This lens directs the beam, whereas the second lens, $L_2$,
images it to a {\em fixed} spot at the prism surface.
A displacement $\Delta a$ leads  to a variation in $\theta$,
given by $\Delta\theta=\Delta a/{n f}$.
The focal length of both lenses is {$f=75~$mm}.
The beam has a minimum waist of {$335~\mu$m} at the surface
($1/e^2$ intensity radius).
We checked the beam collimation and found it almost diffraction limited with
a divergence half-angle of less than {1~mrad}.
We use TM polarization for the EW because it yields a stronger dipole
potential than a TE polarized beam of the same power.
Close to the critical angle, the ratio of potential heights is
approximately $n^2$ \cite{FresnelCoefficients}.

For the EW, an injection-locked single mode laser diode
({\it Hitachi} HL7851G98) provides us with up to {28~mW} of optical power
behind the fiber.
It is seeded by an external grating stabilized diode laser,
locked to the {$^{87}$Rb} hyperfine transition
$5S_{1/2}(F=2)\rightarrow 5P_{3/2}(F'=3)$ of the $D_2$ line ({780 nm}),
with a natural linewidth {$\Gamma=2\pi\times 6.0~$MHz}.
The detuning $\delta$ with respect to this transition defines the optical
potential ${\cal U}_{dip}$ for atoms that are released from the MOT in the
$F=2$ ground state.
We adjust detunings up to {$2\pi\times 500~$MHz} by frequency shifting the
seed beam from resonance, using an acousto-optic modulator.
For larger detunings we unlock the seed laser and set its frequency manually,
according to the reading of an optical spectrum analyzer with {1~GHz}
free spectral range.

Atoms that have bounced on the EW mirror are detected by induced fluorescence
from a pulsed probe beam, resonant with the $F=2\rightarrow F'=3$ transition.
The probe beam travels in the vertical downward direction and has a diameter
of {10~mm}.
The fluorescence is recorded from the side, in the $y$ direction,
by a digital frame-transfer CCD camera ({\it Princeton Instruments}).
The integration time is chosen between {0.1~ms} and {1~ms} and is matched to
the duration of the probe pulse.
Each camera image consists of $400\times 400$ pixels, that were
hardware-binned on the CCD chip in groups of four pixels.
The field of view is {$10.2\times10.2~$mm$^2$} with a spatial resolution of
{$51~\mu$m} per pixel.

A typical timing sequence of the experiment is as follows.
The MOT is loaded from the background vapor during {1~s}.
After {4~ms} of polarization gradient cooling in optical molasses the atoms
are released in the $F=2$ ground state by closing a shutter in the cooling
laser beams.
The image capture is triggered with a variable time delay after releasing
the atoms.
During the entire sequence, the EW laser is permanently on.
In addition, a permanent repumping beam, tuned to the
$5S_{1/2}(F=1)\rightarrow 5P_{1/2}(F'=2)$ transition of the $D_1$ line
({795~nm}), counteracts optical pumping to the $F=1$ ground state by the
probe. We observed no significant influence on the performance of the EW mirror
by the repumping light.

We measure the trajectories of bouncing atoms by taking a series of images with
incremental time delays. A typical series with increments of {10~ms} between
the images is shown in {Fig.\,\ref{fig:TimeSequence}}. Our detection destroys
the atom cloud, so a new sample was prepared for each image. The exposure time
was {0.5~ms}. Each image has been averaged over 10 shots. The fourth image
shows the cloud just before the average bouncing time, {$\bar{t_b}=36.7~$ms},
corresponding to the fall height of {6.6~mm}. In later frames we see the atom
cloud bouncing up from the surface. Close to the prism, the fast vertical
motion causes blurring of the image. Another cause of vertical blur is motion
due to radiation pressure by the probe pulse. The horizontal motion of the
clouds was not affected by the probe. We checked this by comparing with images
taken with considerably shorter probe pulses of {0.1~ms} duration.

\begin{figure}[t]
  \centerline{\epsfxsize=8.0cm\epsffile{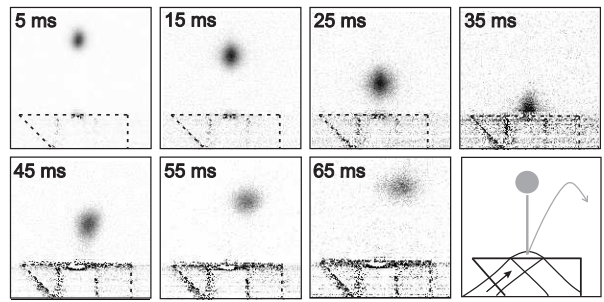}}
  \vspace*{0.5cm}
  \caption{Fluorescence images of a bouncing atom cloud.
	The first image was taken {5~ms} after releasing the atoms from the MOT.
	The configuration of prism and evanescent wave is illustrated by the
	schematic (Field of view: {$10.2\times 10.2~$mm$^2$}).} 
  \label{fig:TimeSequence}
\end{figure}

\subsection{Results}

Radiation pressure in the evanescent wave was observed by analyzing the
horizontal motion of the clouds. From the camera images we determine the center
of mass position of the clouds to about $\pm1$ pixel accuracy.
In {Fig.\,\ref{fig:Trajectories}}, we plot the horizontal position vs. time
elapsed since release from the MOT. We find that the horizontal motion is
uniform before and after the bounce. The horizontal velocity changes suddenly
during the bounce as a consequence of scattering EW photons. The change in
velocity is obtained from a linear fit.

\begin{figure}[thb]
  \centerline{\epsfxsize=9.0cm\epsffile{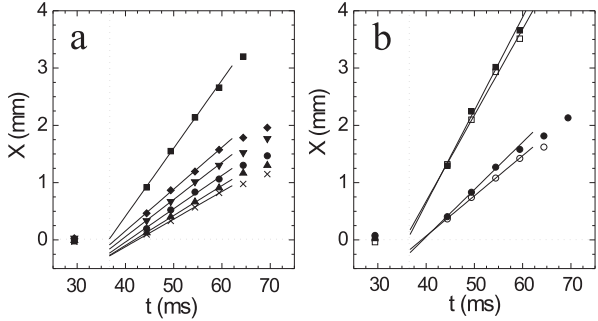}}
  \vspace*{0.5cm}
\caption{Horizontal motion of bouncing atom clouds.
	The center of mass position is plotted vs. time elapsed since release.
	The bounces occur at {36.7~ms} (vertical dotted line). 
    (a)	The EW decay length is varied as $\xi\left(\theta\right)=$ 
	$1.87,~1.03,~0.79,~0.67,~0.59,~0.53~\mu$m
	(from large to small change in velocity).
	The detuning is $44~\Gamma$ and the optical power is {19~mW}.
    (b)	Comparison of two values of EW optical power, {19~mW} (solid points)
	and {10.5~mW} (open points). The detuning is $31~\Gamma$
	and the EW decay lengths are {$\xi=1.87~\mu$m} (large velocity change)
	and {$0.67~\mu$m} (small change). Solid lines indicate linear fits.}
\label{fig:Trajectories}
\end{figure}

\begin{figure}[h]
 \centerline{\epsfxsize=9.0cm\epsffile{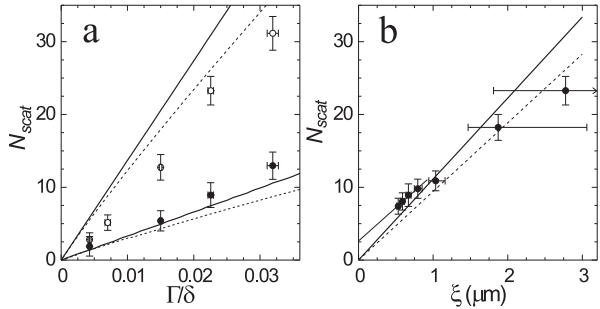}}
 \vspace*{0.5cm}
 \caption{Radiation pressure on bouncing atoms expressed as number of absorbed
	photons, ${\cal N}_{scat}$.
    (a)	Detuning $\delta$ varied for {$\xi=2.8~\mu$m} (open points) and
	{$0.67~\mu$m} (solid points).
    (b)	EW decay length $\xi$ varied for $\delta =44~\Gamma$.
	The laser power was {19~mW}. The thin solid line is a linear fit
	through the first four data points.
	Theoretical predictions: two-level atom
	(see Eq.(\ref{eq:AnalyticalSolution}), thick solid lines).
	Rubidium excited-state hyperfine structure and saturation taken into
	account	(dashed lines).}
\label{fig:Recoils}
\end{figure}

In {Fig.\,\ref{fig:Recoils}}, we show how the radiation pressure depends on the
laser detuning $\delta$ and on the angle of incidence $\theta$. The fitted
horizontal velocity change has been expressed in units of the EW photon recoil,
$p_{rec}=\hbar k_0 n\sin\theta$, with {$\hbar k_0/M=5.88~$mm/s}.

In {Fig.\,\ref{fig:Recoils}a}, the detuning is varied from {188-1400~MHz}
($31-233~\Gamma$).
Two sets of data are shown, taken for two different angles,
{$\theta=\theta_c+0.9~$mrad} and {$\theta_c+15.2~$mrad}.
This corresponds with an EW decay length {$\xi(\theta)=2.8~\mu$m} and
{$0.67~\mu$m}, respectively.
We find that the number of scattered photons is inversely proportional to
$\delta$, as expected.
The predictions based on {Eq.\,(\ref{eq:AnalyticalSolution})} are indicated in
the figure (solid lines).

In {Fig.\,\ref{fig:Recoils}b}, the detuning was kept fixed at $44~\Gamma$ and
the angle of incidence was varied between {0.9~mrad} and {24.0~mrad} above the
critical angle $\theta_c$. This leads to a variation of
$\xi\left(\theta\right)$ from {$2.8~\mu$m} to {$0.53~\mu$m}. Here also, we find
a linear dependence on $\xi\left(\theta\right)$. We see clearly, that a steep
optical potential, {i.e.} a small decay length, causes less radiation pressure
than a shallow potential.

The observed radiation pressure ranges from 2 to 31 photon recoils per atom.
Note, that we separate this subtle effect from the faster vertical motion,
in which atoms enter the optical potential with a momentum of
$p_i\simeq 61~p_{rec}$.

In {Fig.\,\ref{fig:Trajectories}b}, we compare trajectories for {$19\pm1~$mW}
and {$10.5\pm0.5~$mW} optical power in the EW.
As expected from {Eq.\,(\ref{eq:AnalyticalSolution})}, there is no significant
difference in horizontal motion. For a decay length of {$\xi=2.78~\mu$m}
({$0.67~\mu$m}) {\em both} power settings lead to essentially the same
radiation pressure, that is $25\pm3$ ($13\pm2$) scattered photons for {19~mW},
and $23\pm2$ ($11\pm1$) photons for {10.5~mW}. The optical power only
determines the effective mirror surface and thus the fraction of bouncing
atoms. This is also visible in the horizontal width of bouncing clouds.
The bouncing fraction scales with the intensity and the detuning as
$\propto\ln (I/\delta)$. We observe typical fractions of {13\,\%} for
$\delta=44~\Gamma$. For a given optical power, there is an upper limit for the
detuning, above which no bounce can occur.
For the data in {Fig.\,\ref{fig:Recoils}a}, the threshold is calculated as
{$\delta_{th}=6.5~$GHz} ({8.1 GHz}) for {$\xi=2.8~\mu$m} ({$0.67~\mu$m}).
Another threshold condition is implied by the van der Waals interaction,
and yields a lower limit for the EW decay length $\xi$.
For {Fig.\,\ref{fig:Recoils}b} this lower limit is calculated as
{$\xi_{th}=116~$nm}, {i.e.} {$\theta_{th}=(\theta_c+0.59~$rad$)$}.

\subsection{Systematic errors and discussion}

According to {Eq.~(\ref{eq:AnalyticalSolution})}, the radiation pressure
should be inversely proportional to both $\delta$ and $\kappa(\theta)$.
As shown in {Fig.\,\ref{fig:Recoils}}, we find deviations from this
expectation in our experiment, particularly in the $\kappa$ dependence.
A linear fit to the data for {$\xi<1~\mu$m}, extrapolates to an offset of
approximately 3 photon recoils in the limit $\xi\rightarrow 0$
(thin solid line in {Fig.\,\ref{fig:Recoils}b}). The vertical error bars on the
data include statistical errors in the velocity determination from the cloud
trajectories as well as systematic errors. We discuss several possible
systematic errors, namely
(i) the geometric alignment,
(ii) the EW beam angle calibration and collimation,
(iii) diffusely scattered light,
(iv) the van der Waals atom-surface interaction,
(v) excited state contributions to the optical potential,
(vi) and saturation effects.

(i) Geometrical misalignments give rise to systematic errors in the radiation
pressure measurements. For example, a tilt of the prism causes a horizontal
velocity change even for specularly reflected atoms. We checked the prism
alignment and found it tilted {$12\pm5~$mrad} from horizontal. This corresponds
to an offset of $1.5\pm0.6$ recoils on $N_{scat}$. In addition, the atoms are
``launched'' from the MOT with a small initial horizontal velocity which we
found to correspond to less than $\pm0.4$ recoils for all our data.
From {Fig.\,\ref{fig:Trajectories}}, we see that the extrapolated trajectories
at the bouncing time $\bar{t_b}$ do not start from the horizontal position
before the bounce. We attribute this to a horizontal misalignment of the MOT
with respect to the EW spot. Obviously, there is a small displacement of the EW
spot at the prism surface, when adjusting $\theta$ by means of the lens $L_1$
(see {Fig.\,\ref{fig:SetupFigure}b}).
Since the finite sized EW mirror reflects only part of the thermally expanding
atom cloud, such a displacement selects a nonzero horizontal velocity for
bouncing atoms. We corrected for those alignment effects in the radiation
pressure data of {Fig.\,\ref{fig:Recoils}}.
For small radiation pressure values, the systematic error due to
alignment is the dominant contribution in the vertical error bar.

(ii) The uncertainty in the EW angle with respect to the critical angle is
expressed by the horizontal error bars. We determined $\theta-\theta_c$ within
{$\pm0.2~$mrad} by monitoring the optical power transmitted through the prism
surface, while tuning the angle $\theta$ from below to above $\theta_c$. Close
to the critical angle, the decay length $\xi(\theta)$ diverges, and thus the
error bar on $\xi$ becomes very large.
Also the diffraction-limited divergence of the EW beam may become significant.
It causes part of the optical power to propagate into the vacuum. In addition,
the optical potential is governed by a whole distribution of decay lengths.
Thus the model of a simple exponential optical potential
$\propto\exp{\left(-2\kappa z\right)}$ might not be valid and cause the
disagreement of our data with the prediction by
{Eq.\,(\ref{eq:AnalyticalSolution})}. 
For larger angles, {i.e.} {$\xi(\theta)<1~\mu$m}, the effect of the beam
divergence is negligible. This we could verify by numerical analysis.

(iii) Light from the EW can diffusely scatter and propagate into the vacuum
due to roughness of the prism surface. We presume this is the reason for the
extrapolated offset of 3 photon recoils in the radiation pressure
({Fig.\,\ref{fig:Recoils}b}). A preferential light scattering in the direction
of the EW propagating component can be explained, if the power spectrum of the
surface roughness is narrow compared to $1/\lambda$ \cite{HenMolKaiA97}.
The effect of surface roughness on bouncing atoms has previously been observed
\cite{LanLabHen96} as a broadening of atom clouds by the roughness of the
{\em dipole potential}.
In our case, we observe a change in center of mass motion of the clouds
due to an increase in the {\em spontaneous scattering force}.
Such a contribution to the radiation pressure due to surface roughness
vanishes in the limit of large
detuning $\delta$. Thus, we find no significant offset in
{Fig.\,\ref{fig:Recoils}a}. Scattered light might also be the reason for the
small difference in radiation pressure for the two distinct EW power settings,
shown in {Fig.\,\ref{fig:Trajectories}b}. Lower optical power implies slightly
less radiation pressure.

(iv) As stated above, the van der Waals interaction ``softens'' the mirror
potential.
This makes bouncing atoms move longer in the light field,
thus enhancing photon scattering.
We investigated this numerically by integrating the scattering rate along the
atoms' path, including the van der Waals contribution to the mirror potential.
Even for our shortest decay parameter of $0.53~\mu$m, the number of scattered
photons would increase only about {0.8\,\%} compared with
{Eq.\,(\ref{eq:AnalyticalSolution})}, which we cannot resolve experimentally
\cite{vdWaalsSample}.

(v) In a two-level atom model, the scattering rate can be expressed
in the dipole potential as $\Gamma'=(\Gamma/{\hbar\delta})\,{\cal U}_{dip}$.
This is no longer true if we take into account the excited state manifold
$F'=\{0,1,2,3\}$ of {$^{87}$Rb}.
Beside $F'=3$, also $F'=2$ contributes significantly to the mirror potential,
whereas it does not much affect the scattering rate.
With an EW detuning of $44\,\Gamma$, this results in a number of scattered
photons, $N_{scat}$, typically $9\,\%$ lower than expected for a {two-level}
atom, {Eq.\,(\ref{eq:AnalyticalSolution})}.
Here we averaged over contributions from distinct magnetic sublevels.

(vi) In order to investigate the influence of saturation on the number of
scattered photons, $N_{scat}$, we solved the optical Bloch equations
numerically. A bouncing atom encounters the EW as a fast varying light pulse
$I(t)\propto~$sech$^2(\kappa p_i t/M)$, with a typical duration between 3 and
{$10~\mu$s}. By integrating the time-dependent scattering rate for an atom
bouncing with an EW detuning of $44\,\Gamma$, we find
approximately $7\,\%$ less scattered photons compared with the unsaturated
expression of {Eq.\,(\ref{eq:AnalyticalSolution})}.
Note, that the bounces occur sufficiently slow to preserve adiabaticity.
In {Fig.\,\ref{fig:Recoils}}, we show predicted curves,
corrected for hyperfine structure and saturation (dashed solid lines).

\section{Conclusions}

We have directly observed radiation pressure that is exerted on rubidium atoms
while bouncing on an evanescent-wave atom mirror. We did so by analyzing the
bouncing trajectories. The radiation pressure is directed parallel to the
propagating component of the EW, {i.e.} parallel to the interface.
We observe 2 to 31 photon recoils per atom
per bounce. We find the radiation pressure to be independent of the optical
power in the EW, as expected from the exponential character of the EW.

The inverse proportionality to both the EW detuning and the angle of incidence
is in reasonable agreement with a simple two-level atom calculation, using
steady state expressions for the EW optical potential and the photon scattering
rate. The agreement improved when also the excited state hyperfine structure
and saturation effects were taken into account. The measured number of photon
recoils as a function of decay length $\xi$ indicates an offset of approximately
3 recoils in the limit of a very steep EW potential. We assume this is due to
light that is diffusely scattered due to roughness of the prism surface, but
retains a preferential forward direction parallel with the EW propagating
component.

With sufficient resolution, it should be possible to resolve the discrete
nature of the number of photon recoils and also their magnitude,
$\hbar k_x>\hbar k_0$ \cite{MatTakHir98}. 
Our technique could also be used to observe quantum electrodynamical effects
for atoms in the vicinity of a surface, such as radiation pressure out of the
direction of the propagating EW component \cite{HenCou98}.

\section{Acknowledgments}

We wish to thank E.C. Schilder for help with the experiments. 
This work is part of the research program of the ``Stichting voor Fundamenteel
Onderzoek van de Materie'' (Foundation for the Fundamental Research on Matter)
and was made possible by financial support from the ``Nederlandse Organisatie
voor Wetenschappelijk Onderzoek'' (Netherlands Organization for the Advancement
of Research). R.S. has been financially supported by the Royal Netherlands
Academy of Arts and Sciences.

\bibliographystyle{prsty}

\begin{thebibliography}{10}
\bibitem{FresnelCoefficients}
  The enhancement factor of the EW ``intensity'' is calculated for TM and TE
  polarization as
  $T_{TM}={4 n \cos^2\theta\,(2 n^2 \sin^2\theta-1)}
          /{(\cos^2\theta+n^2(n^2 \sin^2\theta-1))}$ and
  $T_{TE}={4 n \cos^2\theta}/{(n^2-1)}$.
\bibitem{vdWaalsSample}
  As an example where van der Waals interaction is significant: For
  {$\delta=1~$GHz}, {$\xi=370~$nm}, and {2.5~mW} optical power, the calculated
  number of scattered photons per bounce increases from 1.09 to 1.13 due to the
  van der Waals interaction.
\end{thebibliography}

\end{document}